\documentclass[sigconf,nonacm]{acmart}

\usepackage{multicol}
\usepackage{hyperref}

\AtBeginDocument{%
	\providecommand\BibTeX{{%
			\normalfont B\kern-0.5em{\scshape i\kern-0.25em b}\kern-0.8em\TeX}}}

\copyrightyear{2019}
\acmYear{2019}
\acmDOI{}

\newcommand{\todo}[1]{\noindent{\normalfont\color{red}\textbf{[}#1\textbf{]}}}
\newcommand{\todonext}[1]{}

\hypersetup{draft}

\begin{document}

\title[]{Peer Reviewing Revisited:\\Assessing Research with Interlinked Semantic Comments}


\author{Cristina-Iulia Bucur}
\affiliation{
	\institution{Vrije Universiteit Amsterdam}
	\city{Amsterdam}
	\country{The Netherlands}}
\email{c.i.bucur@vu.nl}

\author{Tobias Kuhn}
\affiliation{
	\institution{Vrije Universiteit Amsterdam}
	\city{Amsterdam}
	\country{The Netherlands}}
\email{t.kuhn@vu.nl}

\author{Davide Ceolin}
\affiliation{
	\institution{Centrum Wiskunde \& Informatica}
	\city{Amsterdam}
	\country{The Netherlands}}
\email{davide.ceolin@cwi.nl}

\renewcommand{\shortauthors}{}

\begin{abstract}
Scientific publishing seems to be at a turning point. Its paradigm has stayed basically the same for 300 years but is now challenged by the increasing volume of articles that makes it very hard for scientists to stay up to date in their respective fields. In fact, many have pointed out serious flaws of current scientific publishing practices, including the lack of accuracy and efficiency of the reviewing process.
To address some of these problems, we apply here the general principles of the Web and the Semantic Web to scientific publishing, focusing on the reviewing process. We want to determine if a fine-grained model of the scientific publishing workflow can help us make the reviewing processes better organized and more accurate, by ensuring that review comments are created with formal links and semantics from the start. Our contributions include a novel model called Linkflows that allows for such detailed and semantically rich representations of reviews and the reviewing processes. We evaluate our approach on a manually curated dataset from several recent Computer Science journals and conferences that come with open peer reviews. We gathered ground-truth data by contacting the original reviewers and asking them to categorize their own review comments according to our model. Comparing this ground truth to answers provided by model experts, peers, and automated techniques confirms that our approach of formally capturing the reviewers' intentions from the start prevents substantial discrepancies compared to when this information is later extracted from the plain-text comments.  In general, our analysis shows that our model is well understood and easy to apply, and it revealed the semantic properties of such review comments.
\end{abstract}


\maketitle

\section{Introduction}

Scientific articles and the peer reviewing process they undergo are at the core of how scientific research and discoveries are communicated and assessed. For the last 300 years, such scientific articles have played a crucial role in facilitating scientific progress, and peer reviewing has been crucial to ensure the quality and integrity of these scientific contributions. While communication has changed dramatically in the digital age in almost all fields of life, the paradigm of scientific publishing has remained remarkably stable. Journals, articles, and peer reviews are now mostly produced and consumed in digital form, but their structure and appearance has mostly stayed the same.

Although alternative ways of scientific publishing have been explored, most of the suggested approaches still depend on large bulks of text in natural language, digitized and maybe semantically enriched, but essentially following the same old publishing paradigm. While this might not be seen as a problem by itself, the general statistics of scientific publishing point to a serious crisis of information overload \cite{Landhuis2016}. An increasing number of articles is published every day, and the number of publishing scientists worldwide increases with approximately 4--5\% per year \cite{Ware2015}. As a direct consequence, scientists need more time to stay up to date with the recent developments in their respective fields. Already in 2004, for example, epidemiologists in primary care needed more than 20 hours per week if they wanted to read all the new articles in their field \cite{Alper2004}. 


Moreover, we are also facing challenges with respect to the quality of research. Peer reviewing has been seen as a key pillar of the scientific publishing system and the main method of quality assessment. However, despite its established and central role, it is a hotly debated topic in the scientific community due to issues like taking unreasonable amounts of time, lack of transparency, inefficiency, lack of software integration, lack of attribution for reviewers, unsystematic quality assessment, and even poor scientific rigor \cite{Bohannon2013}. Actually, the agreement level between peer reviewers has been found to be only slightly better than what one would expect by chance alone \cite{Lock1985}.

We can tackle some of these problems by applying the ideas and technologies of the Web and the Semantic Web, allowing for a transparent and accessible medium where contributions have a precise structure, can be reliably linked, and can be correctly attributed. In the research presented here, we propose to apply the general principles of the Web and Semantic Web to the reviewing process at a finer-grained scale. These principles entail the use of dereferenceable HTTP URIs to identify, lookup and access resources on the Web, and to interlink them, thereby forming a large network of interconnected resources (or ``things''). Or, in Tim Berners-Lee's words \cite{Bizer2009}, forming ``a web of data that can be processed directly and indirectly by machines" and humans alike. In such a Web-based system, which does not depend on a central authority, everybody can say anything about anything. To put all these statements into context, we can moreover capture provenance information that will allow us to judge the quality and reliability of the provided information. Therefore, following these principles, we can represent information in the form of immutable nodes in a Web-wide distributed network and accurately track the history and provenance of each snippet of information.

This approach of semantically modeled nodes in a network of information snippets can then also use reasoning techniques to aggregate information in dynamic views, instead of the static representations of classical journal articles. Peer reviews can then be represented on a finer-grained scale, with explicit links to the specific parts (e.g. paragraphs) of the scientific contribution they are about. This in turn enables more detailed and more precise quality assessments of both, the scientific contributions and the reviews themselves.


\section{Related work} \label{related-work}

In the context of the Semantic Web, considerable research has been done in formally enriching the meaning of traditional articles. This can facilitate the discoverability of the article and its linking to other related articles or concepts. Important outcomes of these efforts include the SPAR ontologies\footnote{\url{http://www.sparontologies.net/}}, which are a set of core ontologies around semantic publishing, PRISM (the Publishing Requirements for Industry Standard Metadata)\footnote{\url{https://www.idealliance.org/prism-metadata}}, which uses metadata terms to describe published works, and SKOS (the Simple Knowledge Organization System)\footnote{\url{https://www.w3.org/2004/02/skos/}}, which provides a formal backbone model for knowledge organization systems. 

In order to address the information overload crisis, solutions to automatically extract structured information from scientific texts have been proposed. The most common of these include text mining techniques like association rule mining and Inductive Logic Programming powered by large knowledge bases \cite{Kabir2013}. 
These approaches, however, do not attempt to transcend the current publishing paradigm and are therefore limited in their potential to address the core of the problem, which is that knowledge is ``burried'' in narrative texts in the first place, from which it has to be ``mined'' again in a costly and error-prone process \cite{Mons2005}.
To address this problem at its source, alternative approaches have been proposed to make scientific communication more structured and machine-understandable from the beginning.
Efforts in this direction include the concept and technology of nanopublications, which allow for representing scientific claims in small linked data packages based on RDF and formal provenance \cite{Groth2010}, with extensions that allow for informal and partly formal representations to extend their application range  \cite{Kuhn2013}. The micropublication model is a related effort that emphasizes the importance of the argumentation structure when applying formal knowledge representation methods to scientific communication \cite{Clark2014}. Approaches that build upon these concepts include work that combines the micropublication ontology with Open Annotations Data Model to model knowledge about drug-drug interactions and to link them to their scientific evidence in articles \cite{Schneider2014}.
All these approaches are aligned with the general ideas of ``genuine semantic publishing'' \cite{Kuhn2017} and Linked Science \cite{Kauppinen2016}.

With respect to quality assessment of scientific publications, one of the most widely used indicators in the last 40 years is the Journal Impact Factor (JIF)  \cite{Garfield2006, Lariviere2018}, but this metric has been the subject of extensive debates, as it was shown that it can be manipulated  \cite{Kiesslich2016}, and has problems like skewness of citations, false precision,
absence of confidence intervals, and asymmetry in its calculation \cite{Copiello2019}. Also, the JIF can be biased towards journals that publish a larger number of non-research items (e.g. research notes or comments) or publish more articles overall  \cite{Editorial2011}. To address these issues, approaches like the Dataset Quality Information (daQ) \cite{Debattista2014} use Semantic Web technology to enable more accurate and more flexible measures of quality, while the dimensions of quality have been investigated in their own right in the specific context of Linked Data and the Semantic Web \cite{Zaveri2015}. 

A deciding factor for the publication of a scientific article is the evaluation by the peers in the field: the peer reviews. While this system for evaluating the quality of research has been used for a long time, researchers have also outlined the flaws of this process \cite{Smith1988}, advocating for a change. Some point out that there is no conclusive evidence in favor of peer reviews \cite{Linkov2006}, that the existing evidence actually corroborates flaws in this system \cite{Smith2010}, and that biases and preconceptions in judgments affect the overall evaluation of quality that the peer reviewers make \cite{Benda2011}. Proposals that have been put forward to increase the quality of peer reviews, like better review structure \cite{Sucato2018, Leung2014} or paying peer reviewers for their work \cite{Diamandis2017}, have seen little uptake so far. Interesting developments have emerged recently around making reviews more fine-grained \cite{Sadeghi2019} and representing the structure of reviews with a dedicated ontology\footnote{\url{http://fairreviews.linkeddata.es/def/core/index.html}}.

As datasets, documents and knowledge in general are spread on the Web, where everything can be shared, reused, and linked, decentralization is a key concept. Decentralization implies that no central authority --- like a publishing house --- has global control over the content or participants of the system. A large amount of research has investigated the prerequisites and consequences of decentralization in general \cite{Shadbolt2006}, and to a lesser extent specifically in the context of scientific publishing. The former includes recent initiatives like Solid\footnote{\url{https://solid.inrupt.com/}}, a completely decentralized Linked Data framework. The latter include approaches to ensure the functioning of a secure and decentralized global file system \cite{Mazieres1998}, the application of the BitTorrent peer-to-peer file sharing protocol to distribute and access scientific data \cite{Markman2014}, decentralized data publishing approaches for RDF data \cite{Filiali2011}, nanopublications \cite{Kuhn2016}, RASH\footnote{\url{https://github.com/essepuntato/rash}} and Dokieli \cite{Capadisli2017}.



\section{Approach} \label{approach}

In this work, we aim to answer the following research question:
\textit{Can an approach for scientific publishing based on a fine-grained semantic model help to make reviewing better structured and more accurate?}

\subsection{General Approach} \label{general-approach}


In our approach, we apply the general principles of the Web and the Semantic Web to the field of scientific publishing. While the Web consists of a distributed network of documents that link to each other, the Semantic Web adds to that a more fine-grained network at the level of data and knowledge, where the nodes are concepts, i.e. domain entities, not documents.
Following the principles of the Web and the Semantic Web, we argue for fine-grained publication of scientific knowledge in the form of small distributed knowledge snippets that form the nodes in a network, to replace the long prose texts we currently find in articles and reviews. These knowledge snippets are represented with formal semantics, and identified and located with the help of dereferenceable URIs. In this way, scientific knowledge can be shared and accessed as a continuously growing decentralized network of machine-readable snippets of scientific contributions, instead of a static and machine-unfriendly collection of long texts.

\todonext{Figure \ref{fig:review} shows how the peer reviewing part of scientific communication, on which we focus in this work. The figure shows how this reviewing could work with our proposed approach, based on a real example of a review comments that address a paper snippet (from our dataset to be introduced below).
All the interactions between reviewers and authors together with the changes of the initial text snippet can be easily tracked, interpreted, and aggregated. In this way, peer reviews become more transparent and specialized \todo{?}, while we can also develop in the future metrics that can not only evaluate the quality of the "article" snippets, but also the quality of the review itself.}

\subsection{Linkflows Model of Reviewing} \label{linkflows-model}

\begin{figure}[tbp]
	\centering
	\includegraphics[width=\linewidth]{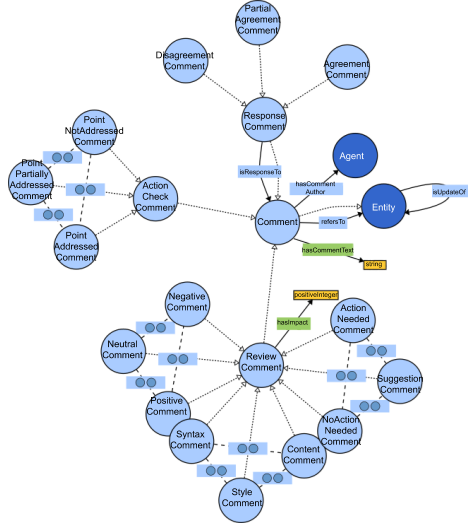}
	\caption{Linkflows model for reviewing. \textmd{Visualization with the online WebVOWL tool \protect\footnotemark.}}
	\label{fig:model}
\end{figure}

At the core of our approach, we propose an ontology for granular and semantic reviewing. Figure \ref{fig:model} shows the main classes and properties of this ontology that we call the Linkflows model of reviewing. The formal ontology specification can be found online.\footnote{\url{https://github.com/LaraHack/linkflows_model/blob/master/Linkflows.ttl}} 

\footnotetext{\url{http://www.visualdataweb.de/webvowl/\#iri=https://raw.githubusercontent.com/LaraHack/linkflows_model/master/Linkflows.ttl}}

The main class of the ontology is the \textit{Comment} class, which includes review comments (subclass \textit{ReviewComment}), on which we focus here, but the class also includes general text annotations or any kind of comment about a text snippet that comes with a dereferenceable URI.
The general properties of the model are \textit{refersTo}, which connects a comment to the entity the comment is about, \textit{isResponseTo}, which declares that a comment is a response to another comment, \textit{isUpdateOf}, which connects an entity such as a comment to its previous version, \textit{hasCommentText}, which links to the text content of a comment, and \textit{hasCommentAuthor}, which declares the person who wrote a comment.

\begin{figure}[tbp]
	\centering
	\includegraphics[width=\linewidth]{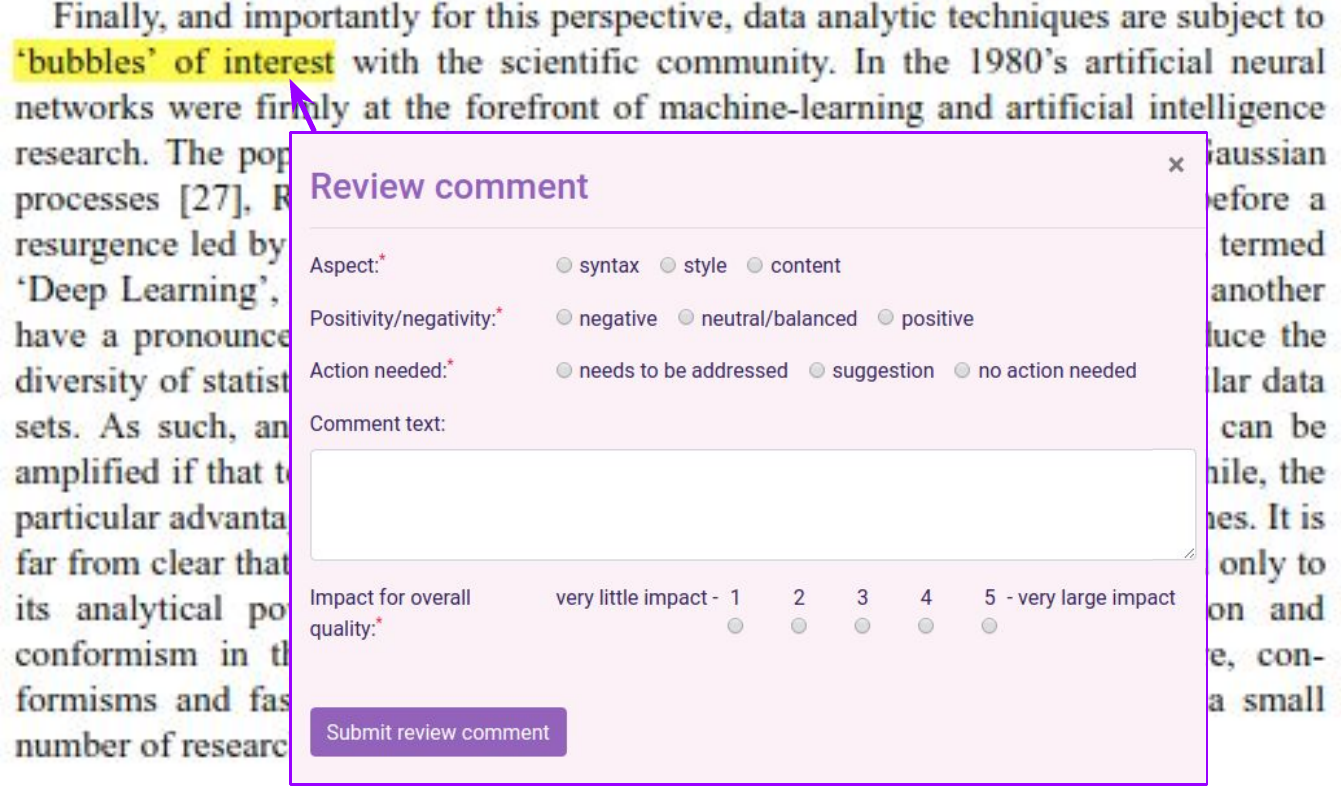}
	\caption{Mock interface implementing the Linkflows model for reviewing.}
	\label{fig:form}
\end{figure}

Our model defines three dimensions for review comments with different categories, each defined as subclasses of the class \textit{ReviewComment}, which form the core of our semantic representation of reviews. The first dimension is about whether the point raised in the review comment is about individual spelling or grammar issues (\textit{SyntaxComment}), the general style of the text including text structure, text flow, text ordering, and consistent wording (\textit{StyleComment}), or the content of the text, e.g. about missing literature, the presented arguments, or the validity of the findings (\textit{ContentComment}). The second dimension is the positivity/negativity of the review comment: \textit{PositiveComment} for review comments that mainly raise positive points, \textit{NeutralComment} for neutral or balanced points raised, and \textit{NegativeComment} for the cases with mainly negative points. The third dimension captures whether an action is needed in response to the review comment (according to the reviewer): \textit{ActionNeededComment} means that the reviewer thinks his or her comment necessarily needs to be addressed by the author; \textit{SuggestionComment} stands for comments that may or may not be addressed; and \textit{NoActionNeededComment} represents the comments that do not need to be addressed, such as plain observations. On top of that, we define a datatype property \textit{hasImpact} that takes an integer from 1 to 5 to represent the extent of the impact of the point raised in the review comment on the overall quality of the article according to the reviewer. For negative comments this score indicates what would be the positive impact if the point is fully addressed, while for positive points it indicates what would be the negative impact if this point were not true.
To further clarify these dimensions and how they could be implemented in an actual reviewing system, we show a mockup of such an interface in Figure \ref{fig:form}.

For representing the interaction between reviewers and authors and follow-up actions that are necessary or requested by the reviewer, the \textit{ResponseComment} and \textit{ActionCheckComment} were created as subclasses of \textit{Comment}.
The author of the text snippet that was commented upon can agree, disagree, or partially agree with the review comment of the reviewer and they can indicate this classifying their response comment accordingly as \textit{AgreementComment}, \textit{PartialAgreementComment}, or \textit{DisagreementComment}.

Finally, reviewers or editors can indicate in another follow-up comment whether they think that the author indeed addressed the point raised by the reviewer (to which they might have agreed, disagreed, or partially agreed). This can be expressed by the subclasses \textit{PointAddressedComment}, \textit{PointPartiallyAddressedComment}, and \textit{PointNotAddressedComment}.

\section{Evaluation Design} \label{evaluation-design}

We evaluate our approach based on recently published articles that come with open peer reviews. We use these data to simulate how the Linkflows model for reviewing would have worked, had this fine-grained semantic model been applied to these manuscripts from the start.
Specifically, we performed the following steps:
First, we created a dataset consisting of 35 articles with open reviews and rebuttal letters (where available).
Second, the review snippets were manually annotated by model experts by applying our model.
Third, we asked reviewers of the selected articles to rate one of their own review comments with regard to our model, thereby giving us ground-truth data.
Fourth, we asked experts via a questionnaire to apply our model to the review snippets for which we have ground-truth answers.
Fifth and lastly, we applied different automated sentiment analysis methods on the review comments to check if these automated methods would be able to correctly identify the positivity or negativity in these review comments. These five steps are explained in more detail below.

\subsection{Dataset} \label{dataset-collection}

In order to study how well our model can capture the reviewing process, we needed a dataset of manuscripts and their reviews. For that, we selected journals and conferences in Computer Science that make their reviews openly available in a non-anonymous way. Because this data preparation involves a lot of manual work, we had to restrict ourselves to a subset of all available articles, and a subset of the reviewers they had.

Specifically, we started by considering  all the 38 articles published in 2018 in the Semantic Web Journal (SemWeb)\footnote{\url{http://www.semantic-web-journal.net/issues\#2018}}, the 13 articles published in the first edition of the journal Data Science (DS)\footnote{\url{https://content-iospress-com.vu-nl.idm.oclc.org/journals/data-science/1/1-2}}, and the 25 articles published in 2018 in the PeerJ in Computer Science Journal (PJCS)\footnote{\url{https://peerj.com/computer-science/}}. Additionally, we collected data from two conferences where article versions and reviews are openly accessible via the openreview.net platform \cite{Soergel2013}. The only conferences in openreview.net that had a complete number of submissions (all articles and reviews) uploaded on the platform at the time we created our dataset were two workshops in 2018: Decentralizing the Semantic Web from the International Semantic Web Conference (ISWC-DeSemWeb)\footnote{\url{https://openreview.net/group?id=swsa.semanticweb.org/ISWC/2018/DeSemWeb\&mode=edit}} and the International Workshop on Reading Music Systems from the International Society for Music Retrieval Conference (ISMIR-WoRMS)\footnote{\url{https://openreview.net/group?id=ISMIR.net}}. ISWC-DeSemWeb had 10 submissions and ISMIR-WoRMS had a total of 12 submissions.

From this set of articles, we first filtered out the ones that were eventually rejected (we had to focus on accepted papers, because the reviews are typically not openly available for rejected ones), had only anonymous reviewers (because of the ground truth step below), or where one of the authors of this paper was editor or reviewer (for objectivity reasons).
After this first filtering step, we selected seven articles from each of the five data sources (the three journals and the two workshops), resulting in a dataset of 35 articles. This selection was done randomly by applying the SHA-256 hash function on their title strings and picking the seven articles with the smallest hash values. We then applied the same procedure on the reviewer names to select a random non-anonymous peer review for each of the selected papers. We always picked the first-round reviews for the papers that went over multiple rounds of reviews.

Finally, we took the resulting reviews and split them into smaller, finer-grained snippets of review comments according to our model. Each of these snippets covers a single point raised by the reviewer, and they often correspond to an individual paragraph or an individual list item of the original review. We did an initial annotation of the review comments corresponding to the 35 selected articles and reviews with regard to the target of the review comments. This target of a review comment can be (a) the entire article, (b) a section or (c) a paragraph (or a similar type of structure in terms of the granularity of the ideas expressed like figure, table, diagram, sentence, footnote, listing, reference, etc.). The level of granularity of the review comments can later indicate, for example, at which level the text received most comments, or to what extent the authors chose to address the issues raised on the different levels. All this is aligned with the Linkflows view of scientific communication as a network of inter-connected distributed information snippets. A review comment and the paragraph it addresses, for example, are two connected nodes in this network, and so are all the paragraphs and sections of an article.


In summary, therefore, our dataset covers 35 pairs of articles and reviews, where the reviews are split into smaller snippets that allow us now to further simulate the application of our model by manual annotation. This dataset is accessible online\footnote{\url{https://github.com/LaraHack/linkflows_reviewing_study}}.

\subsection{Ground-Truth and Manual Annotation}

In order to find out how well the Linkflows model would have worked had it been applied to the papers of our dataset in the first place, we approached the specific peer reviewers who wrote the reviews in our dataset (which is why it was important to exclude anonymous reviewers) and asked them in a questionnaire to categorize one of their existing review comments according to our new model. We can then use their responses as ground-truth data. Specifically, they were asked whether a review comment was about syntax, style or content, whether they raised a positive or negative point, whether an action was mandatory or suggested, what the impact level was for the overall quality of the article, and whether the author eventually addressed the point.

With our envisaged approach, the various actors (most importantly the reviewers) would in the future directly contribute to the network of semantically represented snippets. This kind of annotation activity is therefore only needed for our evaluation methodology but would not be needed in the future when our approach is applied.

Considering the limited time people have for filling in questionnaires, we selected just one review comment for each peer reviewer. For that, we again applied SHA-256 hashing on the review text and chose the one with the smallest hash value. For simplicity, we only considered paragraph-level comments here. To optimize the questionnaire that asks reviewers to provide the classifications according to our model, we conducted a small pilot study in the Semantic Web groups at the Vrije Universiteit Amsterdam. We received 12 answers and incorporated the feedback in the further design of the questionnaire before we sent it to the peer reviewers. Out of 35 contacted peer reviewers (corresponding to the 35 selected articles in our dataset), eleven replied. This therefore resulted in a ground-truth dataset of eleven review comments.

In order to further assess the applicability of the model, the three authors of this paper independently annotated the review comments of the ground-truth datasets as well, considering all review snippets of all reviewers for these articles. This will allow us to compare our answers to the ones of the ground truth, but also to quantify the agreement among us model experts by using Fleiss' Kappa, which can be seen as an indication of the clarity and quality of the model.

For comparison, we also calculate a random baseline of equally distributed and fully uncorrelated responses that correspond to the outcome of random classification. This will serve as an upper baseline for the disagreement with the ground truth.

\todonext{In order to also have a lower baseline, the manual annotation contained a last task that was about reclassifying a random subset of the paragraph-level review comments. This will allow us to quantify ``self-disagreement'', i.e. to what extent an annotator contradict himself/herself when doing exactly the same annotation task twice. \todo{TK: some more details about the number of re-annotated snippets}}

\subsection{Peer Questionnaire} \label{peer-study}

Next, we wanted to find out how important it is to let reviewers themselves semantically express their reviews, instead of extracting this representation afterwards from classical plain-text based reviews. Specifically, we wanted to find out to what extent the interpretation of a review comment by a peer researcher would differ from the meaning that the reviewer had in mind. We hypothesize that there is a substantial level of disagreement, which would demonstrate that we cannot reliably reconstruct the reviewer's intention in detail if we don't let him or her express these explicitly when authoring the review. This in turn, can then be seen an indication of the benefit and value of our approach.

For this part, we used the same kind of questionnaire that we also used for the ground-truth collection, now covering all eleven review snippets that are covered by the ground-truth data. We first conducted again a small pilot study, to see how long it will take to fill in such a questionnaire. Due to the assumed time constraints of potential respondents, we decided to split the questionnaire randomly in two parts, such that filling in one part of the questionnaire would take somewhere between 10--15 minutes. We then sent the two parts of the questionnaire to various mailing lists in the field of Computer Science.

\subsection{Automated Sentiment Analysis} \label{auto-SA}

Finally, we wanted to compare the results we get from human experts with what we can achieve with fully automated methods. Specifically, we applied off-the-shelf sentiment analysis tools on our review texts and compared their resulting sentiment scores to the positivity/negativity values as annotated by the original reviewers.
Sentiment, however, is not exactly the same as positivity/negativity: Sentiment analysis focuses on 'how' things are expressed (i.e. whether positive/polite language is used) while our positivity/negativity dimension captures 'what' is said (i.e. whether the raised point is a positive one). Reviews can be very politely phrased yet contain very serious criticism, thus expressing a negative point with positive or neutral language. Moreover, sentiment analysis tools are typically trained on Web documents such as news items and not on scientific texts or reviews, which have a different vocabulary and language use. We therefore hypothesize that the performance of sentiment analysis tools to detect positivity/negativity in our sense will in general not be satisfactory.

We used a benchmark sentiment analysis platform called iFeel\footnote{\url{http://blackbird.dcc.ufmg.br:1210/}} \cite{Ribeiro2016} to automatically detect whether the review comments in our ground-truth dataset express negative, neutral/balanced, or positive points. We ran all the 18 lexicon-based sentiment analysis methods available of the benchmark platform and collected the results. Finally, we calculated the accuracy of these automated sentiment analysis methods versus the ground-truth data.

\section{Results} \label{results}

We can now have a look at the results we obtained by conducting the studies outlined above.

\subsection{Descriptive Analysis} \label{descriptive-analysis}

\begin{figure}[tbp]
	\centering
	\includegraphics[width=\linewidth]{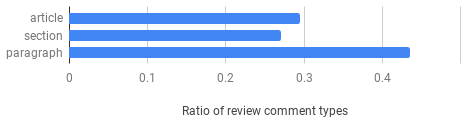}
	\caption{The part of the article that a review comment targets.}
	\label{fig:part}
\end{figure}

Figure \ref{fig:part} shows the distribution of the total of 421 review comments in our dataset with respect to the type of article structure they target. 29\% of them are about the article as a whole, 27\% are at the section level, and the remaining 44\% target a paragraph or an even smaller part of the article. This indicates that the review comments indeed often work at a granular level and that the application of a finer-grained model, such as the Linkflows model, therefore indeed seems appropriate and valuable.

\begin{figure}[tbp]
	\centering
	\includegraphics[width=\linewidth]{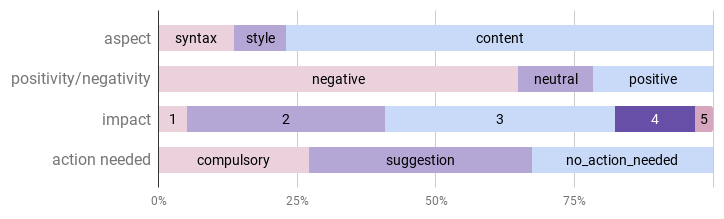}
	\caption{The results of the model expert annotations.}
	\label{fig:manual-annotation}
\end{figure}

Figure \ref{fig:manual-annotation} shows the results of the manual annotation by us model experts on the 450 review comments that correspond to the articles included in the ground-truth subset (articles for which we have one review snippet annotated by one of its peer reviewers). We notice that most of the review comments are about the content of the text, while comments about style and syntax were less common. In terms of the positivity/negativity, we see that --- unsurprisingly --- most of the review comments were rated as negative. The impact of the review commentx for the overall quality of the article is rarely assigned the extreme values of 1 or 5, but the remaining three values of 2--4 are all common (2 and 3 more so than 4). The action that needs to be taken according to the reviewer, finally, shows a balanced distribution of the three categories (compulsory, suggestion, and no-action-needed).


For all of the 450 review snippets that we rated, the average degree of agreement based on Fleiss' kappa \cite{Fleiss1971} had a value of 0.42 which indicates moderate agreement between raters for all dimensions of the Linkflows model \cite{Landis1977}. We notice, however, a considerable variation across dimensions: the aspect and positivity/negativity dimensions had the highest agreement with values of 0.68 and 0.62 respectively (substantial agreement), the action needed had a value of 0.37 (fair agreement), while the impact dimension gave a value of just 0.03 (slight agreement). This low value for the impact dimension is not so surprising given that it consists of just a numerical scale without precise definitions of the individual categories. Moreover, it has a larger number of categories which is known to lead to worse kappa values. But, altogether, the substantial inter-annotator agreement shows that our model is well applicable and sufficiently precise.

\begin{table*}[tbp]
	\caption{Percentages of questions where it was not possible to answer (Peers experiment).}
	\label{tab:no-answer}
	\small
	\begin{tabular}{l|c|c|c|c|c|c|}
		& aspect & positivity/negativity & action needed & impact & action taken & Overall \\
		\hline
		``more context needed'' & 0.75\% & 1.50\% & 2.64\% & 4.90\% & 2.64\% & 2.49\% \\
		``confusing'' & 1.89\% & 1.70\% & 1.32\% & 2.07\% & 0.94\% & 1.58\% \\
		\hline
		Total & 2.64\% & 3.20\% & 3.96\% & 6.97\% & 2.58\% & 4.07\% \\
	\end{tabular}
\end{table*}


For the peer questionnaire study, we gathered 79 responses in total: 43 responses for the first part of the questionnaire containing 5 review comments (215 answers in total) and 36 answers for the second part with 6 review comments (216 answers in total). The participants had to answer a few questions about their background, which revealed that 58.2\% of the respondents have a university degree and that most of them are working in academia (89.9\%). The majority (75.9\%) has advanced knowledge of Computer Science, while only 5\% consider themselves beginners in this field. This confirms that they can indeed be seen as peers of the reviewers and authors of our dataset of Computer Science papers.

When answering the questions, these peers could also choose the options ``More context would be needed; it is not possible to answer'' or ``The review comment is confusing; it is not possible to answer'', when they didn't feel confident to give an answer. We can now take this as an indication of how well the model worked out from their point of view. Table \ref{tab:no-answer} shows these results. We see some variation across the different dimensions, but at a very low level. The two options combined were chosen in less than 7\% of the cases for any of the dimension, and overall in just a bit over 4\% of the cases was one of these two options chosen. This indicates that the dimensions of the Linkflows model are overall very well understood and easy to apply.

\subsection{Accuracy of Sentiment Analysis Methods}

\begin{figure}[tbp]
	\centering
	\includegraphics[width=\linewidth]{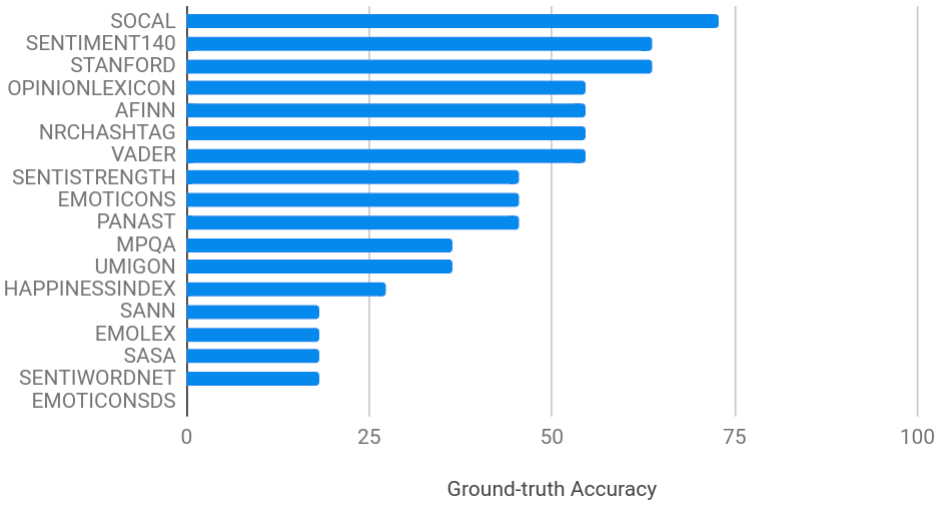}
	\caption{Accuracy of sentiment analysis methods.}
	\label{fig:SA}
\end{figure}

Before we move on to compare the results from the different sources, we first analyze the performance of the different sentiment analysis tools here, in order to be able to chose the best ones for inclusion in the comparative analysis described below.

The results we get from the iFeel platform are already normalized to the values of negative, neutral, and positive, which directly map to the positivity/negativity dimension of our model. We then define accuracy in this context as the ratio of correctly classified instances when compared to the ground truth we collected from the reviewers.

In Figure \ref{fig:SA} we summarized the results of the 18 sentiment analysis methods used to rate the eleven review comments for which we have ground-truth data. We see a wide variation of performance with a maximum accuracy of 72.8\% for the SOCAL method \cite{Taboada2011}. As expected, most of these methods perform quite poorly. In general, we can observe that the sentiment analysis methods that have more complex rules perform best, even if their lexicon size is not large. Apart from SOCAL, we also select the tools ranked second and third for our comparative analysis below: Sentiment140 \cite{Go2009} and Stanford Recursive Deep Model \cite{Socher2013} (both 63.6\% accuracy).


\subsection{Comparative Analysis}

\begin{table*}[tbp]
	\caption{Disagreement scores for four of the Linkflows model dimensions.}
	\label{tab:comparative-aspect}
	\small
	\begin{tabular}{l|ccc|ccc|ccc|ccc|}
		& \multicolumn{3}{c|}{aspect} & \multicolumn{3}{c|}{action needed} & \multicolumn{3}{c|}{impact} & \multicolumn{3}{c|}{action taken}\\
		& R & M & P & R & M & P & R & M & P & R & M & P\\
		\hline
		Reviewer (R) & 0 & & & 0 & & & 0 & & & 0 & & \\
		Model experts (M) & 0.32 & 0 & & 0.26 & 0 & & 0.21 & 0 & & 0.38 & 0 &\\
		Peers (P) & 0.29 & 0.12 & 0 & 0.29 & 0.17 & 0 & 0.31 & 0.14 & 0 & 0.34 & 0.21 & 0\\
		\hline
		\hline
		Random baseline & & 0.67 & & & 0.67 & & & 0.50 & & & 0.67 & \\
	\end{tabular}
\end{table*}

\begin{table}[tbp]
	\caption{Disagreement scores for the positivity/negativity dimension.}
	\label{tab:comparative-pos}
	\small
	\begin{tabular}{l|cccccc|}
		& R & M & P & SA1 & SA2 & SA3\\
		\hline
		Reviewer (R) & 0 & & & & & \\
		Model experts (M) & 0.32 & 0 & & & & \\
		Peers (P) & 0.16 & 0.24 & 0 & & & \\
		SOCAL (SA1) & 0.26 & 0.41 & 0.24 & 0 & & \\
		SENTIMENT140 (SA2) & 0.30 & 0.16 & 0.22 & 0.34 & 0 & \\
		STANFORD (SA3) & 0.30 & 0.16 & 0.22 & 0.34 & 0 & 0 \\
		\hline
		\hline
		Random baseline & \multicolumn{6}{c|}{0.58} \\
	\end{tabular}
\end{table}

We can now have a closer look into the differences between the answers we got from the different sources, to better understand the nature and extent of the observed disagreement.
We quantify this disagreement by applying a variation of the Mean Squared Error metric. To compare the responses of two groups, we calculate normalized (0--1) within-group averages of their responses and then take the square root of the mean squared differences between the groups. For the nominal dimensions, we calculate the squared differences from the ratio for each category separately.

These disagreement scores between reviewers, model experts, and peers for all dimensions of the model are shown in Table \ref{tab:comparative-pos} (for the positivity/negativity dimension, which includes the sentiment analysis tools) and Table \ref{tab:comparative-aspect} (for all other dimensions).

Looking at Tables \ref{tab:comparative-aspect} and \ref{tab:comparative-pos} we see that the agreements between the reviewers and the groups of model experts and peers range from 0.12 to 0.38, well above perfect agreement (0) but also well below the random baseline (0.5 for the ordinal dimension \textit{impact}, 0.58 for the ordinal dimension \textit{positivity/negativity}, and 0.67 for the nominal dimensions). It is notable that the model experts and the peers always agree with each other more than they agree with the ground truth in the form of the original reviewer. This seems to indicate that they misinterpret the review comments in a relatively small but consistent manner. The highest disagreement scores for all pairs of groups come from the \textit{action taken} dimension, which seems in general slightly more difficult than the others.

With respect to the positivity/negativity dimension (Table \ref{tab:comparative-pos}), we see that the best automated sentiment analysis tool surprisingly performed a bit better than the model experts (0.26 versus 0.32 disagreement score, compared to the ground truth). A further surprising outcome is that the peers performed much better than the model experts and the sentiment analysis tools. We will further investigate this effect below.
Moreover we see that the second and third best sentiment analysis tools provide identical results (and therefore zero disagrement), but that they have considerable disagreement with the best tool (SOCAL), which might hint at complementary information given by these tools and therefore the potential for an ensemble method.

Table \ref{tab:experts-peers} focuses on the comparison of the two main groups of model experts and peers. We see that for three of the five dimensions, the peers had a higher agreement (i.e. lower disagreement) with the ground truth than the model experts. Testing these observed differences for statistical significance with a two-tailed Wilcoxon signed-rank test, we see however that none of these differences is significant.

From the somewhat surprising result that model experts do not perform significantly better than peers, we could conclude that expertise on the model doesn't actually help with the task. It could also be that the benefit of this expertise was offset by the fact that the model experts shared more detailed information and acquired background knowledge through in-depth discussions with each other, and therefore they had information neither the reviewers nor the peers had. This might have made the reviewers and the peers settle on similar choices that are different from the model experts'. Because the group of peers consists of more than ten times as many individuals as compared to the model expert group, an alternative explanation is that a wisdom of the crowd effect is involved, where averaging over a larger number of individual responses gets us closer to the truth.

\begin{table*}[tbp]
	\caption{Model experts versus peers.}
	\label{tab:experts-peers}
	\small
	\begin{tabular}{l|r|r|r|r|r|}
		& aspect & positivity/negativity & action needed & impact & action taken\\
		\hline
		Model experts & 0.32 & 0.32 & 0.26 & 0.21 & 0.38\\
		Peers & 0.29 & 0.16 & 0.29 & 0.31 & 0.34\\
		\hline
		Difference & 0.03 & 0.16 & --0.03 & --0.10 & 0.04 \\
		$p$-value (Wilcoxon signed-rank test) & 0.21 & 0.28 & 0.42 & 0.09 & 0.66\\
		\hline
		Average of groups of 3 peers & 0.34 & 0.22 & 0.31 & 0.34 & 0.44\\
		Standard deviation of groups of 3 peers &  0.026 & 0.010 & 0.119 & 0.055 & 0.016\\
	\end{tabular}
\end{table*}

In order to test this wisdom of the crowd hypothesis, we divided the answers from the peers into twelve groups of three peers each (thereby matching the size of the model expert group) and calculated the disagreement scores for each of these smaller groups. The bottom part of Table \ref{tab:experts-peers} shows the average results. We see that the average disagreement of the smaller peer groups is always larger than the disagreement of the large peer group, confirming that there seems to have been a wisdom of the crowd effect at play. Except for positivity/negativity, where the difference was largest to start with, the average disagreement of the smaller peer groups is larger than for the model experts. Therefore, there seems to be indeed a wisdom of the crowd effect at play that makes larger groups perform better than smaller ones.

\section{Discussion and Conclusion} \label{conclusion}


The results of this study suggest that the Linkflows model is indeed able to express at a finer-grained level the properties of review comments and their relation to the article text. We propose to let reviewers create semantically represented review comments from the start. Our evaluation showed that a substantial level of disagreement arises if other actors like peer researchers or model experts try to reconstruct the intended nature of these review comments afterwards, thereby underlying the importance of capturing this information right at the source. Through a wisdom of the crowd effect, larger groups of peers can achieve lower disagreement with the ground-truth provided by the original reviewers, but the disagreement is still substantial and larger groups of annotators require more collective effort. Existing automated methods like sentiment analysis tools can reach the same level of agreement like model experts, but a substantial level of disagreement remains for both of them. In summary, this confirms the value of our approach to precisely capture the network and type of review comments directly from the reviewers.

While we have provided a simple mock-up of an interface for reviewers, an actual prototype still has to be implemented and evaluated as future work. Specifically, we plan to experiment with applying the Dokieli environment \cite{Capadisli2017} and Linked Data Notifications \cite{Capadisli2017LDN}. To publish reviews (and also the articles themselves) as a network of semantically annotated snippets in a decentralized way, we will furthermore investigate the application of the nanopublication technology and infrastructure \cite{Kuhn2016} for reliable semantic publishing.



\begin{acks}
The authors thank IOS Press and the Netherlands Institute for Sound and Vision, who partly funded this research.
\end{acks}

\bibliographystyle{ACM-Reference-Format}
\bibliography{base.bib}


	
\end{document}